\documentclass[twoside,slac_one]{revtex4}
\usepackage{aas_macros}
\usepackage{graphicx}
\usepackage{fancyhdr}
\usepackage{amsmath} 
\usepackage{bm}
\usepackage{amsxtra}
\usepackage{amssymb}
\usepackage{amsthm}
\usepackage{latexsym}
\usepackage{lscape}
\usepackage{natbib}

\begin{document}

\title{Ultra High Energy Cosmology with POLARBEAR}

%

\author{B. Keating, S. Moyerman, D. Boettger, J. Edwards, G. Fuller, F. Matsuda, N. Miller,  H. Paar, G. Rebeiz, I. Schanning, M. Shimon, N. Stebor}
\affiliation{Center for Astrophysics and Space Sciences, University of California at San Diego, La Jolla, CA USA}
\author{K. Arnold}
\author{D. Flanigan}
\author{W. Holzapfel}
\author{J. Howard} 
\author{Z. Kermish}
\author{A. Lee} 
\author{M. Lungu} 
\author{M. Myers} 
\author{H. Nishino}
\author{R. O'Brient} 
\author{E. Quealy} 
\author{C. Reichardt}
\author{P. Richards}
\author{C. Shimmin} 
\author{B. Steinbach} 
\author{A. Suzuki}
\author{O. Zahn}
\affiliation{Department of Physics, University of California at Berkeley, Berkeley, CA USA}
\author{J. Borrill}
\author{C. Cantalupo}
\author{E. Kisner}
\author{E. Linder}
\author{M. Sholl}
\author{H. Spieler}
\affiliation{Physics Division, Lawrence Berkeley National Laboratory, Berkeley, CA 94720}
\author{A. Anthony}
\author{N. Halverson} 
\affiliation{Department of Astrophysical and Planetary Sciences, University of Colorado}
\author{J. Errard}
\author{G. Fabbian}
\author{M. Le Jeune}
\author{R. Stompor}
\affiliation{Astroparticule et Cosmologie, CNRS, Universit{\'e} Paris-Diderot, Paris, France}
\author{A. Jaffe}
\author{D. O'Dea}
\affiliation{Department of Physics, Imperial College}
\author{Y. Chinone}
\author{M. Hasegawa}
\author{M. Hazumi}
\author{T. Matsumura}
\author{H. Morii}
\author{A. Shimizu}
\author{T. Tomaru}
\affiliation{High Energy Accelerator Research Organization (KEK), Tsukuba, Ibaraki, Japan}
\author{P. Hyland}
\author{M. Dobbs}
\affiliation{Physics Department, McGill University}
\author{P. Ade}
\author{W. Grainger}
\author{C. Tucker}
\affiliation{School of Physics and Astronomy, University of Cardiff}

\begin{abstract}
Observations of the temperature anisotropy of the Cosmic Microwave Background (CMB) lend support to an inflationary origin of the universe, yet no direct evidence verifying inflation exists. Many current experiments are focussing on the CMB's polarization anisotropy, specifically its curl component (called ``B-mode" polarization), which remains undetected. The inflationary paradigm predicts the existence of a primordial gravitational wave background that imprints a unique B-mode signature on the CMB's polarization at large angular scales. The CMB B-mode signal also encodes gravitational lensing information at smaller angular scales, bearing the imprint of cosmological large scale structures (LSS) which in turn may elucidate the properties of cosmological neutrinos. The quest for detection of these signals; each of which is orders of magnitude smaller than the CMB temperature anisotropy signal, has motivated the development of background-limited detectors with precise control of systematic effects. The POLARBEAR experiment is designed to perform a deep search for the signature of gravitational waves from inflation and to characterize lensing of the CMB by LSS. POLARBEAR is a 3.5 meter ground-based telescope with 3.8 arcminute angular resolution at 150 GHz. At the heart of the POLARBEAR receiver is an array featuring 1274 antenna-coupled superconducting transition edge sensor  (TES) bolometers cooled to 0.25 Kelvin. POLARBEAR is designed to reach a tensor-to-scalar ratio of 0.025 after two years of observation -- more than an order of magnitude improvement over the current best results, which would test physics at energies near the GUT scale. POLARBEAR had an engineering run in the Inyo Mountains of Eastern California in 2010 and will begin observations in the Atacama Desert in Chile in 2011.
\end{abstract}

\maketitle

\thispagestyle{fancy}

\vspace{-10pt}
\section{Introduction}
The discovery of the temperature anisotropy in the Cosmic Microwave Background (CMB) has motivated the development of large, background-limited detector arrays capable of high-precision data. A string of remarkable experiments have mapped the angular power spectrum of the CMB's temperature anisotropy with exquisite precision, constraining many parameters describing the origin of our universe. The emergent model predicts an expanding primordial universe with scalar density perturbations seeding the formation of gravitationally bound structure.

While our understanding of the universe's first few minutes stands as a stunning triumph of the past century, we know almost nothing about its initial conditions. Inflation is a bold cosmological paradigm which predicts that a superluminal expansion of space-time took place a mere $10^{-36}$ seconds after the Big Bang. In this brief, highly-energetic epoch, space-time and gravity were both purely quantum in nature. The quantum gravitational fluctuations, called the Gravitational Wave Background (GWB), produced by Inflation survived until decoupling when they encoded the CMB with a unique polarization pattern at $\simeq 2^\circ$ angular scales called ``B-mode" polarization. If detected, the B-mode signal would be a smoking gun, determining the Inflationary origin of the Universe beyond a reasonable doubt. The CMB also bears the imprint of gravitational lensing at $0.1^\circ$ angular scales, the result of primordial dark matter, including neutrinos, warping space-time and altering CMB polarization. The gravitational lensing CMB signal depends crucially on the masses of primordial neutrinos. Fortuitously, this gravitational lensing polarization signal, arguably equally important to cosmology as the GWB signal, is guaranteed to exist (though it has never been detected). 

The CMB's polarization can be decomposed into two components known as E-modes and the above-mentioned B-modes, titled as such analogously to electric and magnetic fields. Acoustic oscillations provide only even-parity E-mode polarization, so any B-mode signal must be generated by different physics or phenomena subsequent to last-scattering ~\cite{PhysRevLett.78.2058, PhysRevLett.78.2054}. Whereas the E-mode signal has been detected by several experiments and agrees well with the model of acoustic oscillations, the B-mode remains undetected to date ~\cite{chiang2010, Bischoff:2010ud}.  At arcminute angular scales, the structure of B-mode polarization arises from gravitational lensing of the CMB's primordial E-mode polarization. As photons from the last scattering surface free-stream, the E-mode polarization signal is lensed by gravitational interactions into a parity violating pattern with B-mode (anti) symmetry ~\cite{0004-637X-574-2-566}. In the absence of gravitational lensing, the CMB's B-mode signal at small angular scales would be negligible, making gravitational lensing a very clean probe of the lensing signal. The gravitational lensing structure serves as the earliest probe of the formation of gravitationally bound structures and allows for constraints on the sum of the neutrino masses ~\cite{PhysRevLett.91.241301}.  

At larger angular scales, B-mode polarization generated by primordial gravitational waves may exist. As these gravitational waves propagate through the early universe, they stress and shear the surface of last-scattering, imprinting a B-mode signal. If detected, any B-mode polarization due to gravitational waves would provide robust evidence of an early inflationary epoch and allow characterization of the energy scale at which inflation took place. This energy scale is characterized by the parameter $r$,  the ratio of tensor-to-scalar perturbations, $E \simeq 10^{16} \rm{GeV} \Big{(}\frac{r}{0.01}\Big{)}^{\frac{1}{4}}$. Thus, detecting the large angular scale B-mode polarization pattern will provide a probe of fundamental physics at energy scales \emph{one trillion times} higher than any earthbound particle accelerators! To detect both B-mode polarization signals requires sensitive detectors, high angular resolution, and superb control over systematic effects. POLARBEAR has been designed to meet these stringent goals.
\vspace{-10pt}
\section{POLARBEAR Science Goals}

\begin{figure}[ht]
\centering
\includegraphics[width=120mm]{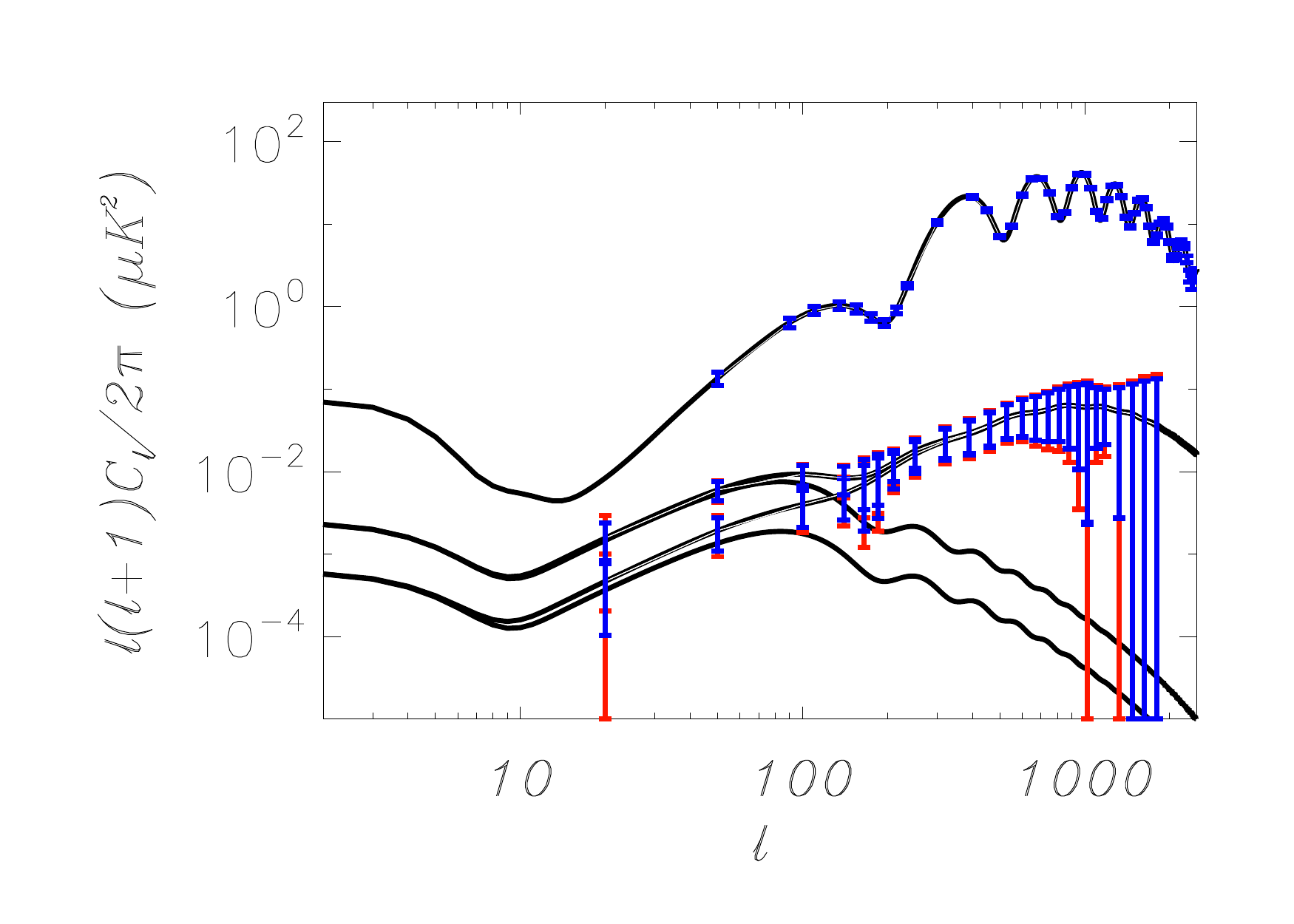}
\caption{Projection of POLARBEAR limits on the B-mode power spectra after one observing season. The inflation model plots with $r$=0.1 and $r$ = 0.025 are also shown for comparison.} \label{Figure 1}
\end{figure}

Figure 1 shows the projection of the POLARBEAR CMB angular power spectrum for both E-mode and B-mode polarization. The projection is for a total observation time of 8 hours per day for 9 months. The blue error bars are modeled with no foreground consideration, and red error bars accounts for foregrounds following  ~\cite{Bowden:2003ub, Sethi:1998xx, tegmark-1999}. Two B-mode gravitational wave spectra are shown for comparison: $r = 0.1$ and 0.025. POLARBEAR is projected to achieve a 2$\sigma$ detection of $r = 0.025$ and will also characterize the gravitational lensing signals in an attempt to measure properties of cosmological neutrinos ~\cite{Smith:2008an, 1475-7516-2010-05-037}.

\vspace{-10pt}
\section{POLARBEAR Design}

POLARBEAR consists of the POLARBEAR receiver and the Huan Tran Telescope (HTT) and will be located at the James Ax Observatory in the Atacama Desert in Chile. POLARBEAR is designed to have unprecedented sensitivity over an extremely large range of angular scales, surveying four $15^\circ \times15^\circ $ regions (to detect the large angular scale B-modes from the GWB) with 3.8 arcminute resolution with precise control and mitigation of systematic effects. The following section describes the instrument's detectors and optics. For more details on the design of POLARBEAR we refer the reader to Arnold et al (2010) \cite{Arnold2010}.

\subsection{Huan Tran Telescope}

The Huan Tran Telescope (Figure 2) is an off-axis Gregorian telescope with a 3.5 meter primary aperture that provides the 3.8 arcminute angular resolution necessary to characterize the gravitational lensing signal. The primary is composed of a 2.5 meter high-precision monolithic reflector surrounded by a 0.5 meter radius guard ring, which is used to prevent beam spillover and minimize sidelobes. Incident radiation is directed from the primary to the baffled secondary mirror that re-images the primary image onto the detector array. The telescope was designed for large optical throughput while simultaneously mitigating systematic effects such as temperature-to-polarization leakage and cross-polarization. The inner shield blocks stray light from entering the optical path  ~\cite{PBsys}. 

\begin{figure}[t]
\centering
\includegraphics[width=70mm]{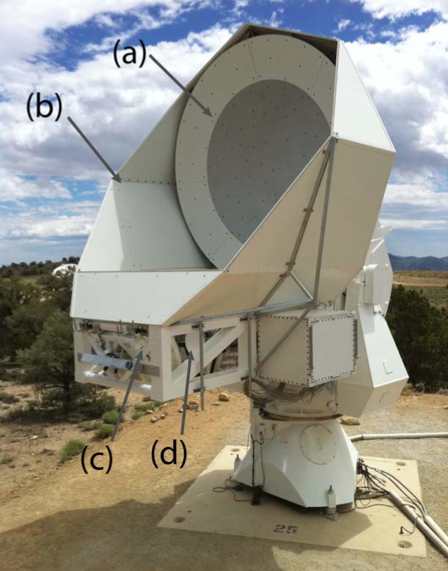}
\caption{Image of the Huan Tran Telescope at the Inyo Mountain site in 2010. The labeled components are (a) the primary mirror, (b) the co-moving inner shield, (c) the baffled secondary mirror, and (d) the cryogenic receiver.} \label{Figure 2}
\end{figure}

\subsection{Receiver Optics}

The fundamental noise limit for any CMB experiment is known as photon noise, which is set by the quantum fluctuations in the arrival rate of photons. CMB polarimeters seek to minimize all other noise sources (\emph{e.g.}, phonon noise, readout noise, \emph{etc}), such that photon noise dominates. To achieve a high level of sensitivity, the POLARBEAR focal plane is cooled to 250 milliKelvin so that thermal carrier noise in the detectors is smaller than the photon noise. POLARBEAR achieves this cooling with closed-cycle refrigeration: a pulse tube cooler and a three-stage helium sorption refrigerator. 

\begin{figure}[b]
\centering
\includegraphics[width=160mm]{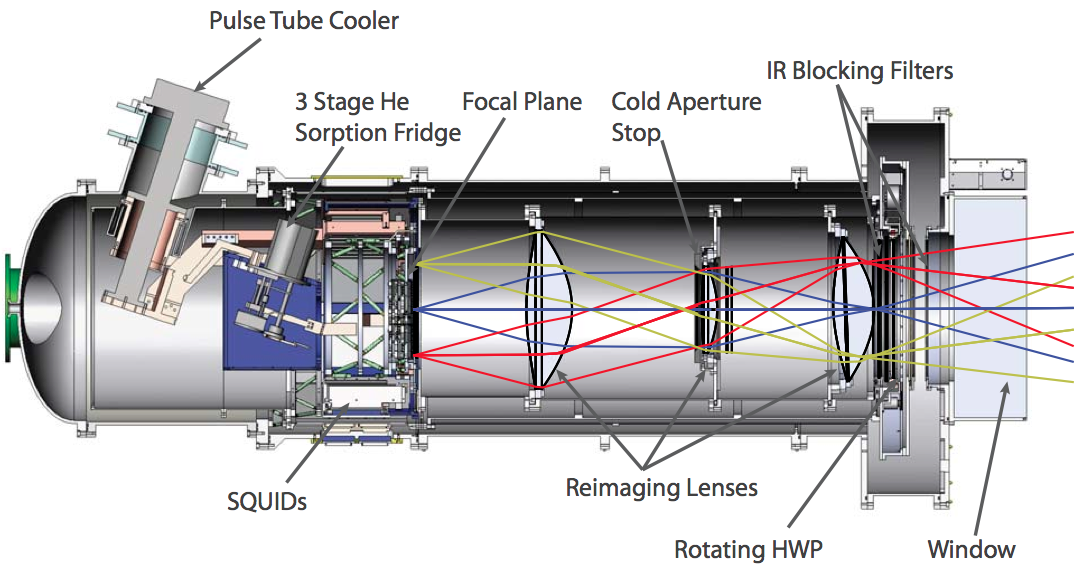}
\caption{Cross section of the POLARBEAR receiver with optical ray trace paths shown.} \label{Figure 3}
\end{figure}

Figure 3 shows a cross-section through the receiver. Before incidence on the focal plane, millimeter-wave radiation entering the receiver encounters, sequentially, a Zotefoam vacuum window, a rotating sapphire half wave plate (HWP), single- and multi-layer metal mesh filters, porous Teflon radiation filters, and anti-reflection coated ultra-high molecular weight polyethylene re-imaging lenses. The re-imaging lenses and multi-layer metal mesh filters are emissive in the spectral band of the detectors and are therefore cooled to prevent an increase in optical loading. 

The rotating half-wave plate (HWP) is a 3.1 mm thick single crystal disk of A-plane sapphire. Because of its birefringent properties, rotation of the HWP modulates the polarization of the signal (only), and thus allows for mitigation of instrumental systematic effects which do not have the requisite symmetry of true CMB polarization. The HWP is cooled to 70K to reduce thermal emission. POLARBEAR's focal plane features a planar array of bolometers, requiring a flat, telecentric image from the primary reflector. This is achieved using three re-imaging lenses coupled to the telescope optics. A cold aperture stop with an absorbing edge images the inner high-precision monolithic section of the primary mirror, suppressing sidelobe response.

\subsection{Detectors and Focal Plane}

While there are a number of competing detector technologies for CMB experiments, only bolometers operated from sub-Kelvin platforms and at frequencies $\geq$ 90 GHz are sensitive enough to be photon noise limited. A detector in this regime cannot make significant gains in sensitivity by improvements to the detector itself, instead, gains are made by increasing the number of detectors and the throughput of the telescope. POLARBEAR is designed to meet this goal, using entirely lithographed superconducting transition edge sensor (TES) bolometers in scalable arrays. Figure 4a shows an image of the entire POLARBEAR focal plane. The focal plane is composed of seven separate hexagonal sub-arrays. One such array, along with readout electronics, is shown in Figure 4b.      

\begin{figure}[t]
\centering
\includegraphics[width=180mm]{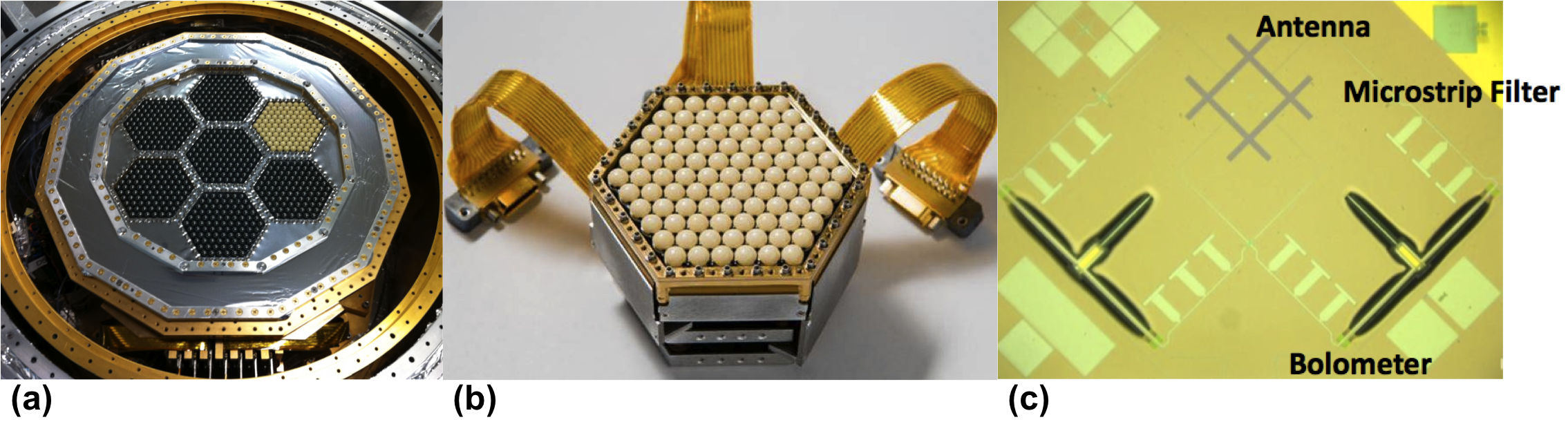}
\caption{(a) An image of the full POLARBEAR focal plane as seen from the top (b) A single hex wafer plus vertically integrated readout (c) Microscope image of a single POLARBEAR pixel with important components labeled} \label{Figure 4}
\end{figure}

The POLARBEAR focal plane features 637 pixels (1274 TES bolometers) with a spectral band between 120 and 180 GHz to avoid atmospheric contamination. Each pixel (Figure 4c) features several lithographed components: a double-slot dipole antenna that separates millimeter-wave signals into orthogonally polarized components, distributed bandpass filters that reject frequencies outside of the desired observing band, and superconducting transition edge sensors that are thermally isolated to provide photon noise limited detection of the CMB ~\cite{MyersAntennaCoupled, MyersFP}. 

Each antenna is coupled to an anti-reflection coated silicon lenslet that serves to increase forward gain, decrease radiation lost to substrate modes, and magnify the effective size of the antenna (thus increasing the area available for the rest of the pixel). The lenslets are quarter-wavelength anti-reflection coated to minimize reflection loss at the surface. The coupling of the extended hemispherical lenslet above the planar antenna produces a diffraction-limited beam with Gaussicity and directivity similar to that of a conical horn ~\cite{slotdipole}. 


Another benefit of superconducting TES bolometers is that they can be read-out and multiplexed by low noise SQUID amplifiers, allowing for simultaneous readout of thousands of detectors. POLARBEAR detector signals are read-out with a frequency-domain multiplexed readout using cryogenic SQUID ammeters. A group of eight transition edge sensor (TES) bolometers is AC voltage-biased, each with a different frequency, and its current response is summed using a single SQUID. The ``frequency comb" SQUID signals exit the cryostat and are demodulated at 300 K. This multiplexed readout reduces the otherwise prohibitive amount of wiring, and is designed as a vertical stack (Figure 4b) allowing POLARBEAR to take full advantage of (precious) cold focal plane area.  
\vspace{-10pt}
\section{Recent Results}
\vspace{-5pt}
An engineering run of POLARBEAR was performed in 2010 at the CARMA site in the Inyo Mountains of Eastern California. The focal plane of this engineering run contained three of seven hexagonal sub-arrays and a 50\% attenuating filter at 4K was placed in the optical path to reduce the high atmospheric power present in California (which will be absent in the James Ax Observatory in the Atacama Desert).  Bright astrophysical point sources were observed in order to characterize POLARBEAR's beam parameters. Data was accumulated for each source by scanning the telescope in azimuth and stepping in elevation. In this way, each pixel crosses the source many times. Co-added maps of the brightest celestial source, Jupiter, are shown in Figure 5a. These maps give a best fit Gaussian beam with full width at half maximum of 3.8 arcminutes.

Observations of the bright, polarized mm-wave source Tau A were performed during the Cedar Flat engineering run. A row of 12 observing pixels were chosen and rastered back and forth across across Tau A for approximately two hours while the HWP was stationary. Maps were subsequently produced by rotating the HWP and reobserving Tau A. Figure 5b shows the co-added temperature and polarization maps of Tau A. The polarization magnitude and angle show good agreement with published results ~\cite{refId0}.

\begin{figure}[b]
\centering
\includegraphics[width=170mm]{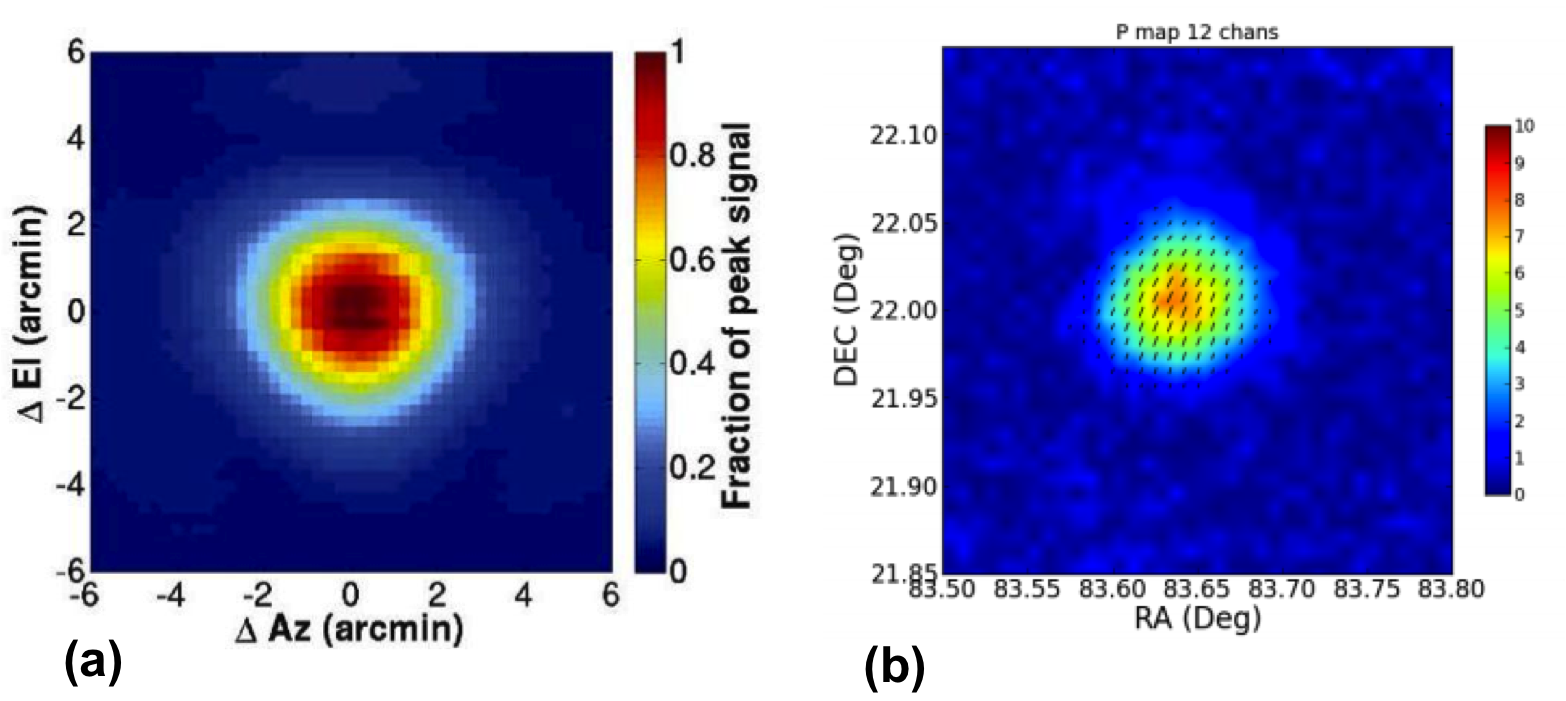}
\caption{(a) Co-added map of Jupiter used to characterize beam gaussicity, ellipticity, and beam-width. b) Temperature and polarization map of Tau-A from 2 hours of observing data with 12 pixels. Data for both plots were obtained during the 2010 Inyo Mountains, California engineering run.} \label{Figure 5}
\end{figure}

Other tests (such as polarization response) were also performed during the Cedar Flat engineering run to characterize beam parameters. POLARBEAR is more than a factor of two better than the most stringent requirements for differential pointing, differential ellipticity, and differential beam size in order to detect CMB B-Modes from inflationary gravitational waves with $r = 0.025$. 
\vspace{-10pt}
\section{Future Plans}
\vspace{-10pt}
The goals of POLARBEAR are to detect the gravity-wave background (GWB) and the imprint of massive neutrinos on the gravitational lensing power spectrum. To this end, the POLARBEAR team is pursuing a long range strategy featuring multiple telescopes, which are currently under construction, coupled to novel wide-band antennas and TES bolometers. Here, we briefly outline our long range strategy. We will not refer to the telecope described previously as POLARBEAR-1 and the long range project as POLARBEAR.
\vspace{-10pt}
\subsection{POLARBEAR-1}
\vspace{-5pt}
The goal of POLARBEAR-1 is the detection of the GWB down to the limit imposed by gravitational lensing (when viewed as a foreground) \cite{0004-637X-574-2-566} which equates to $r\simeq0.025$ at the 2$\sigma$ level and the detection of lensing B-modes. POLARBEAR-1 is currently located in the Atacama Desert in Chile. Initially, POLARBEAR-1 will observe the same sky regions as QUIET has observed at 40 GHz ~\cite{Bischoff:2010ud}, so that the combined data set will have broad multi-frequency coverage. Observing identical regions of the sky at multiple frequencies allows for spectral mitigation of, arguably, the most pernicious systematic effect: millimeter wave foreground emission from the Milky Way galaxy.
\vspace{-10pt}
\subsection{POLARBEAR-2}
\vspace{-5pt}
In 2014, POLARBEAR will be upgraded with a new receiver mounted on an identical 3.5 meter diameter Huan Tran Telescope at the James Ax Observatory; a configuration called ``POLARBEAR-2". POLARBEAR-2 uses extremely broadband ``sinuous'' antennae and will have six times as many bolometers as POLARBEAR-1, allowing us to constrain even lower inflationary energy scales than standard inflationary models. POLARBEAR-2 pixels  feature planar, sinuous antennas with an increased fractional bandwidth (3.5:1 vs 1:3 for the current double slot dipole) with good cross-polarization rejection ~\cite{sinuous,sinuous2}. The high level of cross-polarization rejection will minimize systematic polarization while the wide bandwidth allows for separation of the signal from a single antenna into a number of different frequency bands. Each antenna will feed two bands, 90 and 150 GHz, defined by microstrip filters. Each frequency will have two associated bolometers with orthogonal polarizations, giving the POLARBEAR-2 twice the bolometers per pixel. The design for the POLARBEAR-2 focal plane will feature 1837 pixel pixels (7348 bolometers) and is shown in Figure 6a. POLARBEAR-2 will operate simultaneously with the original HTT and POLARBEAR-1 receiver. The combination of the two telescopes and receivers will be capable of mapping the B-mode power spectrum to the same level of precision as current E-mode measurements. 

\begin{figure}[ht]
\centering
\includegraphics[width=160mm]{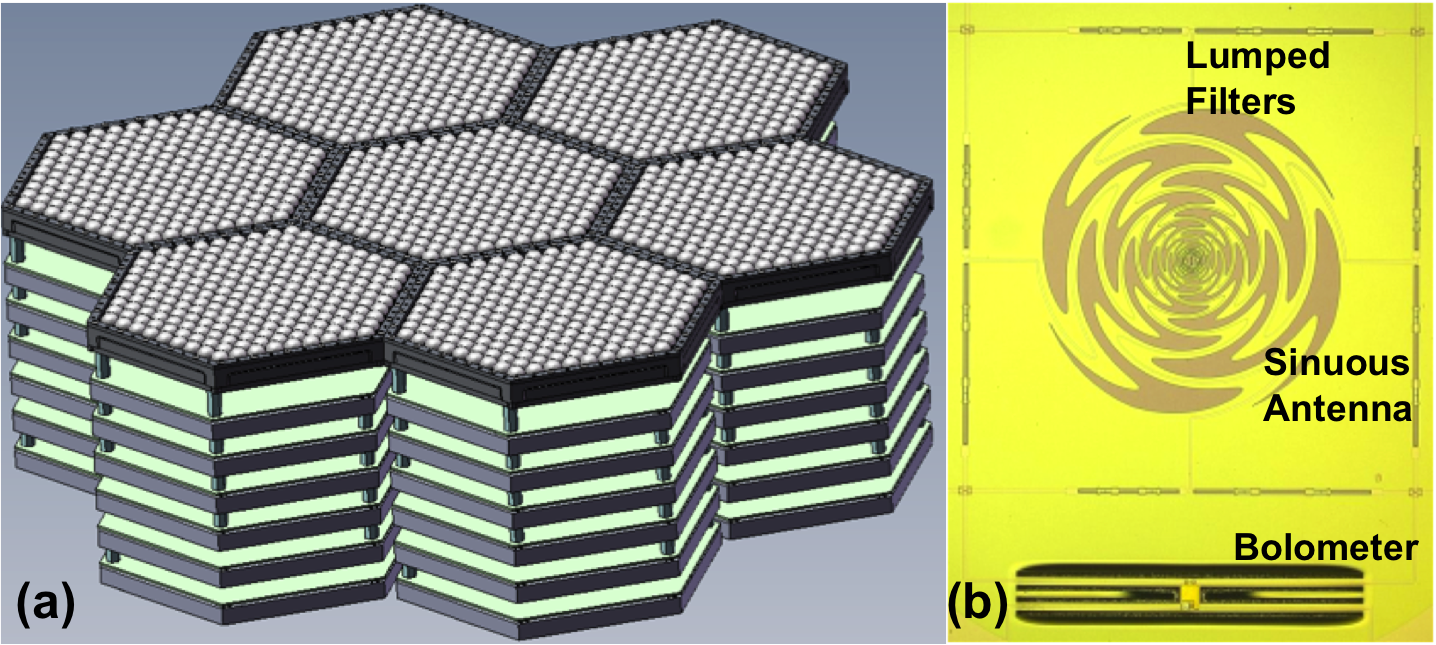}
\caption{(a) Rendering of the POLARBEAR-II focal plane with 1500 pixels (b) POLARBEAR-II prototype pixel with sinuous antenna, lumped filters, and bolometer.} \label{Figure 6}
\end{figure}
\vspace{-10pt}
\subsection{POLARBEAR Extended}
\vspace{-5pt}
POLARBEAR Extended (PBX) is the expansion of the POLARBEAR project to three Huan Tran Telescopes, each equipped with POLARBEAR-2 style detectors at the James Ax Observatory. The dual-band receivers will be operational at 90 and 150 GHz or 150 and 220 GHz. The receiver mounted on the first HTT telescope (POLARBEAR-1) will also be replaced by the multi-frequency POLARBEAR-2 style receiver. Dramatic gains in sensitivity can be achieved with three telescopes outfitted with more than 7000 bolometers each. POLARBEAR Extended will perform a deep search for the GWB and map out the B-mode power spectrum to unprecedented precision. These CMB polarization measurements will constrain the sum of the neutrino masses down to 0.1 eV.
\vspace{-10pt}
\subsection{Future research: Superconducting Tunnel Junction Refrigeration}
\vspace{-5pt}
The POLARBEAR team is also currently investigating the use of a new technology, normal-insulating-superconducting junctions (NIS), that can be lithographed into planar arrays along for on-chip refrigeration. NIS junctions, when voltage biased, exploit quantum mechanical tunneling to remove the hottest electrons from the normal metal and dissipate the heat into the superconductor. The normal metal is then extended onto the suspended bolometer, giving the bolometer an effective bath temperature much lower than the rest of the system. While POLARBEAR is photon limited in the sense that photon statistics are the dominant source of noise, there is still a noise presence intrinsic to detector thermal fluctuations that scales with temperature. NIS junctions have the advantage of continuous cooling and are cheap to fabricate. Groups using similar technology have developed NIS junctions capable of cooling of x-ray sensors from 250 mK down below 160 mK ~\cite{oneil09, silverberg08}, though none have been used for observations to date. A detailed mathematical model has been constructed to evaluate the performance of NIS junctions coupled to the POLARBEAR bolometers. With careful engineering design and an optimal bias, we predict a temperature drop by nearly a factor of two, from 250 mK to 130 mK.  This would improve the raw sensitivity of the experiment by  15\% and the mapping speed by more than 18\%.

\vspace{-10pt}\vspace{-10pt}
\begin{acknowledgments}
\vspace{-10pt}
The POLARBEAR project is funded by the National Science Foundation under grant AST-0618398. Antenna-coupled bolometer development at Berkeley is funded by NASA under grant NNG06GJ08G. The McGill authors acknowledge funding from the Natural Sciences and Engineering Research Council and Canadian Institute for Advanced Research. MD acknowledges support from an Alfred P. Sloan Research Fellowship and Canada Research Chair program. The KEK authors and the POLARBEAR-2 receiver system development are supported by MEXT KAKENHI 21111002. All silicon wafer-based technology is fabricated at the UC Berkeley Microlab. 

\end{acknowledgments}

\bibliographystyle{unsrt}
\bibliography{myrefs}

\end{document}